\documentclass[conference]{IEEEtran}
\IEEEoverridecommandlockouts
\usepackage[margin=0.58in]{geometry}
\usepackage{times,amsmath,amsfonts,amssymb,amsthm}
\usepackage{cite}
\usepackage{graphicx,subfigure,psfrag}
\usepackage{algorithm}%,algorithmic}
\usepackage{booktabs}
\usepackage{multirow}
\usepackage{multicol}
\usepackage{colortbl}
\usepackage{color}
\usepackage{algpseudocode}
\usepackage{balance}
\def\abf{{\bf a}}

\def\dbf{{\bf d}}

\def\pbf{{\bf p}}

\def\Ac{{\cal A}}

\def\Dc{{\cal D}}

\def\Kc{{\cal K}}

\def\Nc{{\cal N}}

\def\Qc{{\cal Q}}

\def\Sc{{\cal S}}
\def\Tc{{\cal T}}
\def\Uc{{\cal U}}

\def\Wc{{\cal W}}

\def\ie{{\it i.e.,\ \/}}
\def\nn{\nonumber}

\def\Unique{\text{Unique}}

\theoremstyle{definition}
\newtheorem{lemma}{Lemma}
\newtheorem{theorem}{Theorem}
\newtheorem{Remark}{Remark}

\newtheorem{myDef}{Definition}

\begin{document}

\title{Memory-Rate Tradeoff for Caching with Uncoded Placement under Nonuniform File Popularity
}
 \author{\IEEEauthorblockN{Yong Deng and Min Dong\\}
 \IEEEauthorblockA{Dept. of Electrical, Computer and Software Engineering, Ontario Tech University, Ontario, Canada
 }}

\maketitle

\begin{abstract}
For caching with nonuniform file popularity, we aim to characterize the memory-rate tradeoff under  uncoded cache placement. We consider the recently proposed Modified Coded Caching Scheme (MCCS) with the optimized cache placement  based on the popularity-first approach to minimize the average delivery rate. We  introduce two information-theoretic lower bounds on the average rate for caching under uncoded placement.
For $K=2$ users, we show that the  optimized MCCS attains the lower bound and is optimal for caching with uncoded placement. For  general $K$ users with distinct file requests, the optimized MCCS attains the popularity-first-based lower bound. When there are  redundant file requests among $K$ users, we show  a possible gap between the optimized MCCS and the lower bounds, which is attributed to zero-padding commonly used for coded delivery.  We analyze  the impact of zero-padding  and its limitation. Simulation study shows that the loss  is  very small in general and only exists in some limited cases.
\end{abstract}\vspace{-1.2em}

\allowdisplaybreaks
%\begin{IEEEkeywords}
%component, formatting, style, styling, insert
%\end{IEEEkeywords}
\section{Introduction}
Caching is expected to play a critical role in future wireless networks to alleviate the growingly intensive data traffic  and ensure timely delivery. %\cite{Bastug&etal:COMMag14}.
A \emph{Coded Caching Scheme} (CCS) has recently been proposed  in  \cite{Maddah-Ali&Niesen:TIT2014} to significantly improve the caching gain. The scheme combines a cache placement scheme for  uncoded content storage and a coded multicasting delivery strategy.
It explores both global and local caching gain to achieve substantial delivery rate (load) reduction. Since then, coded caching has drawn considerable attention, extending to various system models or network scenarios \cite{Niesen&Maddah-Ali:TIT2015,ji2016fundamental,Xu&etal:TIT17}. A \emph{Modified Coded Caching Scheme} (MCCS) has  been proposed in~\cite{Yu&Maddah-Ali:TIT2018} with an improved delivery strategy that results in a further reduced delivery rate than the CCS.
The MCCS has then been applied to the device-to-device  networks~\cite{yapar2019optimality}.

For understanding the fundamental limit of the  coded caching, many research efforts are devoted to characterizing the memory-rate tradeoff  in caching with uncoded placement.
This tradeoff has been extensively studied for files with uniform file popularity by developing an achievable scheme and comparing it to an information-theoretic lower bound. With fewer users than files in the system, it is shown that the CCS with optimized cache placement achieves the exact memory-rate tradeoff for the peak rate consideration~\cite{Wan2016On1,Wan2016On2}.
In general, for both peak and average rates,  the MCCS characterizes the exact memory-rate tradeoff  under uniform file popularity\cite{Yu&Maddah-Ali:TIT2018}.
For the more practical scenario with nonuniform file popularity, there is an unknown gap between the state-of-art achievable rates
~\cite{Niesen&Maddah-Ali:TIT2017,Ji&Order:TIT17,Zhang&Coded:TIT18,Daniel&Yu:TIT19,Jin&Cui:Arxiv2017,Jin&Cui:Arxiv2018,Deng&Dong:arxiv2020,Yong2020Optimal} and the lower bound. Only recently, a coded caching scheme was proposed for the case of two files and was demonstrated to achieve the exact memory-tradeoff under uncoded placement~\cite{Sahraei2019TheOptimal}. In general, characterizing the memory-rate tradeoff   is challenging, and how optimal the MCCS is under nonuniform file popularity remains unknown.

A key issue for the MCCS  under nonuniform file popularity is the cache placement design, \ie how to partition each file into subfiles to be stored in user local caches. For uniform file popularity, the symmetric cache placement (\ie the same placement for all files) is optimal for both the CCS~\cite{Daniel&Yu:TIT19} and the MCCS~\cite{Yu&Maddah-Ali:TIT2018,Deng&Dong:SPAWC19}.
For nonuniform file popularity, the cache placement is asymmetric among files, resulting in nonequal subfile sizes that complicate both  design and analysis.
To reduce the complexity,  heuristic approaches using file grouping  were proposed to impose the symmetric placement within each group~\cite{Niesen&Maddah-Ali:TIT2017,Ji&Order:TIT17,Zhang&Coded:TIT18}.
Among existing works for cache placement optimization~\cite{Daniel&Yu:TIT19,Jin&Cui:Arxiv2017,Jin&Cui:Arxiv2018,Deng&Dong:arxiv2020,Yong2020Optimal}, a   popularity-first strategy that allocates more memory to the more popular file has been shown to be optimal   for  the CCS~\cite{Jin&Cui:Arxiv2017} and numerically verified to be optimal for the MCCS\cite{Jin&Cui:Arxiv2018}.
 For the CCS, the optimal cache placement has recently been  completely characterized, which shows that the file grouping is the inherent structure \cite{Deng&Dong:arxiv2020}. For the MCCS, the optimal placement based on the popularity-first strategy is obtained \cite{Yong2020Optimal}. In the above works, zero-padding  is commonly adopted to simplify the coded delivery. However, its impact on the optimality of the coded delivery is unknown and has never been studied to our best knowledge.

In this paper, we aim to characterize the memory-rate tradeoff for caching under  nonuniform file popularity.
We formulate the cache placement optimization problem for the MCCS under the popularity-first placement, and consider it as the optimized MCCS.
For caching with  uncoded placement, we develop two lower bounds on the average rate: One is for any cache placement. The other is a  lower bound for the popularity-first cache placements only.
We then characterize the memory-rate tradeoff by comparing the average rate of the optimized MCCS with the lower bounds. We prove that, for $K=2$ users, the optimized MCCS is the optimal caching scheme under uncoded placement.
This indicates that  popularity-first placement is optimal, and  zero-padding does not cause any loss of optimality. For $K>2$  users with distinct file requests, we show that the optimized MCCS is the optimal caching scheme under the popularity-first-based uncoded placement, where zero-padding causes no additional loss. When there are redundant file requests among $K>2$ users,  we show that there is a possible gap between the optimized MCCS and the lower bounds. We analyze the loss attributed to zero-padding  and reveal its limitation. Nonetheless, our numerical studies show that the loss is very small  in general and only exists  in very limited cases.

%%%%%%%%%%%%%%%%%%%%%%%%%%%%%%%%%%
\section{System Model }\label{sec:model}

We consider a cache-aided transmission system with a server connecting to $K$ cache-equipped users over a shared error-free link.
 The server has a database consisting of $N$ files $\{W_1,\ldots,W_N\}$. Each file $W_n$ is of size $F$ bits and with  probability  $p_n$ to be requested. Let $\pbf\triangleq[p_1,\ldots,p_N]$ denote the popularity distribution of  the $N$ files, where $\sum_{n=1}^{N}p_n=1$.
We label files according to the decreasing order of their popularities: $p_1\geq p_2\geq \cdots\geq p_N$.
Each user $k$  has a local cache of size $M$, representing cache capacity $MF$ bits,   where  $M$ is a  real number within the interval $[0,N]$.
Denote the file and user index sets by  $\Nc\triangleq\{1,\ldots,N\}$  and $\Kc\triangleq\{1,\ldots,K\}$, respectively.

The coded caching operates in two phases: the cache placement phase and the content delivery phase.
In the cache placement phase, a portion of uncoded file contents from $\{W_1,\ldots,W_N\}$ are placed in each user $k$'s\ local cache, according to a cache placement scheme.
Assume each user $k\in\Kc$ independently requests a file with index $d_k$  from the server.
Let  $\dbf\triangleq[d_1,\ldots,d_K]$ denote the demand vector of $K$ users. In the content delivery phase, based on the demand vector  $\dbf$ and the cached contents at users, the server generates coded messages containing uncached portions of requested files and transmits to the users.
Upon receiving the coded messages, each user $k$  reconstructs its requested file $\hat{W}_{\dbf,k}$ from the received coded messages and its cached content.

\allowdisplaybreaks
%%%%%%%%%%%%%%%%%%%%%%%%%%%%%%%%%%%%%
\section{ Cache Placement Optimization for MCCS}\label{sec:prob}

The MCCS is recently proposed \cite{Yu&Maddah-Ali:TIT2018} as an improvement to the original CCS \cite{Maddah-Ali&Niesen:TIT2014}  to  reduce the delivery rate further. In this section, we formulate the cache placement design for the MCCS  into an optimization problem.

%%%%%%%%%%%%%%%%%%%%%%%%%%%%%%%%%%%%%%%%
\subsubsection{Cache Placement}\label{III.A}

The cache placement of the MCCS is based on file partitioning. For $K$ users, there are total $2^K$ user subsets in $\Kc$, with subset sizes ranging from $0$ to $K$. Among them, there are $\binom{K}{l}$ different user subsets with the same size $l$, for $l=0,\ldots,K$ (including the empty subset of size $0$).
They form a cache subgroup that contains all user subsets of size $l$,  defined as $\Ac^l\triangleq\{\Sc: |\Sc|=l,\ \Sc\subseteq \Kc\}$ with $|\Ac^l|=\binom{K}{l}$, for $l=0,\ldots,K$. Partition each file $W_n$  into $2^K$ non-overlapping subfiles, one for each unique user subset $\Sc\subseteq\Kc $, denoted by
 $W_{n,\Sc}$. Each user in user subset $\Sc$ stores subfile $W_{n,\Sc}$ in its local cache. Note that $W_{n,\Sc}$ can be $\emptyset$, and for $\Sc=\emptyset$,  subfile $W_{n,\emptyset}$  is not stored in any user's cache but only kept   at the server. For any caching scheme, each  file should be able to  be reconstructed  by combining all its subfiles. Thus, we have\vspace{-0.6em}
\begin{align}\label{cons:file_partition_0}
\sum_{l=0}^{K}\sum_{\Sc\in \Ac^l}|W_{n,\Sc}|=F,\quad \text{for}\ n \in \Nc.
\end{align}\\[-2em]

To reduce  the number of variables and simplify the optimization problem for its tractability, we impose the following condition: C1)
 For each file $W_n$, the size of its subfile $W_{n,\Sc}$ only depends on  $|\Sc|$, \ie  $|W_{n,\Sc}|$ is the same for  any $\Sc \in \Ac^l$ of the same size.
It is numerically verified in \cite{Jin&Cui:Arxiv2018} that imposing this condition results in no loss of optimality.
As a result, the subfiles of file $W_n$  are grouped  into file subgroups, each denoted  by $\Wc^l_n=\{W_{n,\Sc}: \Sc \in \Ac^l\}$, for $l=0,\ldots,K$. As a result, there are ${K \choose l}$ subfiles of the same size in  $\Wc^l_n$ (intended for user subsets in cache subgroup $\Ac^l$), and there are total  $K+1$ file subgroups. Following this, let $a_{n,l}$ denote  the  size of subfiles in $\Wc^l_n$, as a fraction of  file $W_n$ size,  \ie $a_{n,l} \triangleq |W_{n,\Sc}|/F$ (for $\forall \Sc \in \Ac^l$),  $l=0,\ldots,K$,   $n\in \Nc$.    Note that $a_{n,0}$ represents the fraction of file $W_n$  that is not stored at any user's cache but only remains at the server. Then,  the file partition constraint \eqref{cons:file_partition_0} is simplified to\vspace{-0.7em}
\begin{align}
  \sum_{l=0}^{K}{K \choose l}a_{n,l}=1, \ n \in \Nc.\label{Constraint1.1}
\end{align}\vspace{-0.8em}

Recall that each subfile is intended for a unique user subset. For the cache placement, user $k$  stores all the subfiles in $\Wc_n^l$ that are  intended for user subsets that contain the user, \ie $\{W_{n,\Sc}: k\in \Sc \text{~and~} \Sc \in \Ac^l\}\subseteq \Wc_n^l$, for  $l=1,\ldots,K$.
Note that in each $\Ac^l$, there are total ${K-1 \choose l-1}$  different user subsets containing the same user $k$. Thus,  there are  $\sum_{l=1}^{K}{K-1 \choose l-1}$ subfiles in each file $W_n$ that a user can  store in its local cache. With subfile size $a_{n,l}$,  this means  that in total, a fraction  $\sum_{l=1}^{K}{K-1 \choose l-1}a_{n,l}$  of file $W_n$ is cached by a user. With  cache size $M$ at each user, we have the following cache  constraint\vspace{-0.5em}
\begin{align}
\sum_{n=1}^{N}\sum_{l=1}^{K}{K-1 \choose l-1}a_{n,l} \leq M.\label{Constraint2}
\end{align}\\[-2em]

For nonuniform file popularity, we consider the popularity-first cache placement approach below. It  simplifies the cache placement optimization problem for coded caching.
\begin{myDef}[Popularity-first placement]\label{def:popfir}
The popularity-first  cache placement is to allocate more cache memory to a more popular file, specified by the following condition
%\begin{align}
$a_{n,l}\ge a_{n+1,l}$, $n\in\Nc\backslash\{N\}, l\in\Kc$.%\label{equ:popu_fir}
%\end{align}
\end{myDef}

\begin{Remark}The popularity-first approach has been used for the cache placement for both the CCS\cite{Daniel&Yu:TIT19,Jin&Cui:Arxiv2017} and the MCCS\cite{Jin&Cui:Arxiv2018,Yong2020Optimal} to simplify the cache placement problem. For the CCS, the popularity-first placement has been proven to be the property of the optimal cache placement\cite{Jin&Cui:Arxiv2017}. For the MCCS, the same is difficult to prove, but it has been verified  numerically  for the optimality of the popularity-first placement \cite{Jin&Cui:Arxiv2018}. In Section \ref{sec:exact}, we will show the optimality of popularity-first  placement for the case of $K=2$ users with arbitrary $N$ files.\end{Remark}

\subsubsection{Content Delivery}\label{subsec:delivery}
In the content delivery phase, the server multicasts coded messages to  different user subsets. Each coded message corresponds to a user subset $\Sc$, formed by bitwise XOR operation  of subfiles as   $C_\Sc\triangleq\bigoplus_{k \in \Sc} \! W_{d_k,\Sc\backslash\{k\}}$.
In the original  CCS, the server simply delivers the coded messages  formed by all the user subsets, for any demand vector  $\dbf$. However, under random demands, multiple users may request the same (popular) file, causing redundant coded messages.
In the MCCS, a modified coded delivery strategy is proposed to remove this   redundancy to  reduce the average delivery rate further. For demand vector $\dbf$, assume that there are $\widetilde{N}(\dbf)$  distinct file requests, where $\widetilde{N}(\dbf)\le K$. To describe the delivery strategy, we first provide the following two definitions:\\
\emph{D1)  Leader group}: The leader group $\Uc$ is a user subset of size   $|\Uc|=\widetilde{N}(\dbf)$, and the users in  $\Uc$ have $\widetilde{N}(\dbf)$ distinct requests.\\
\emph{D2) Redundant group}: Any user subset $\Sc \subseteq \Kc$ with $\Sc \cap \Uc=\emptyset$ is a redundant group;  otherwise, it is a non-redundant group.

In the MCCS, only the coded messages formed by  the non-redundant groups $\{C_\Sc(\dbf): \forall \Sc\subseteq\Kc$ and $\Sc \cap \Uc\neq\emptyset\}$ are  multicasted  to both non-redundant and redundant groups.
With nonuniform file popularity,  subfile partitioning in different files may be different, leading to different subfile sizes.  Subfiles in a coded message are zero-padded to the size of the
largest subfile among them.

\begin{Remark}
Note that zero-padding is a common technique used to form coded messages in the existing works~\cite{Daniel&Yu:TIT19,Jin&Cui:Arxiv2017,Jin&Cui:Arxiv2018,Deng&Dong:arxiv2020,Yong2020Optimal,Saberali&Lampe:TIT20}. However, the impact of zero-padding on the optimality of coded caching has not been studied and is unknown. In Section~\ref{sec:exact}, we will provide our results to answer this  question.
\end{Remark}\vspace{-0.6em}

%%%%%%%%%%%%%%%%%%%%%%%%%%%%%%%%%%%%
\subsubsection{Cache Placement Optimization}

Consider file partition $\{a_{n,l}\}$ for the cache placement and demand vector $\dbf$. Let $\abf_n\triangleq[a_{n,0},\ldots,a_{n,K}]^T$ denote the $(K+1)\times 1$ cache placement vector for file $W_n$, $n\in\Nc$, and let $\abf\triangleq[\abf_1^T,\cdots,\abf_N^T]^T$ represent the entire placement for $N$ files. By zero-padding, for coded message  $C_\Sc$  formed by non-redundant group  $\Sc$ of size $l+1$, its size is  given by $   |C_\Sc|=\max_{k\in\Sc}a_{d_k,l}$.
The delivery rate is the total size of  the coded messages by all the non-redundant groups, given by $R_{\text{MCCS}}(\dbf;\abf)\triangleq\!\sum_{\Sc\subseteq\Kc,\Sc \cap \Uc\neq\emptyset}|C_\Sc|$.
The average delivery rate $\bar R_{\text{MCCS}}$ is given by%\vspace{-0.5em}
\begin{align}\label{avgR_MCCS}
 \!\bar{R}_{\text{MCCS}}(\abf)=\mathbb{E}_\dbf\!\left[R_{\text{MCCS}}(\dbf;\abf)\right]
\!\!=\!\mathbb{E}_\dbf\!\Big[\!\sum_{\Sc\subseteq\Kc,\Sc \cap \Uc\neq\emptyset}\!\!\!\max_{k\in\Sc}a_{d_k,l}\Big]
\end{align}\\[-1em]
where $\mathbb{E}_\dbf[\cdot]$ is taken w.r.t.   $\dbf$.

Define the popularity-first placement set $\Qc\triangleq\{\abf:a_{n,l}\ge a_{n+1,l}, \; \forall n\in\Nc\backslash\{N\}, l\in\Kc\}$. In other words,  $\Qc$ is the set of all popularity-first placements defined in Definition \ref{def:popfir}. The cache placement  optimization problem for the MCCS is then formulated as optimizing  $\abf\in\Qc$ to minimize $\bar{R}_\text{MCCS}$ in \eqref{avgR_MCCS}
\vspace{-0.3em}
\begin{align}
\textrm{\bf P0}: \;\min_{\abf\in\Qc}\;\;  \bar{R}_{\text{MCCS}}(\abf) \quad
 \textrm{s.t.} &\;\;
\eqref{Constraint1.1},\eqref{Constraint2}, \; \text{and~}\nn\\[-.5em]
&\abf_{n}\succcurlyeq\mathbf{0},\; n\in\Nc. \label{Constraint_gt0}
\end{align}\\[-3.5em]

\begin{Remark}
 The optimal cache placement solution  for \textrm{\bf P0} is obtained in \cite{Yong2020Optimal}, where it is shown that the cache placement  under  nonuniform file popularity has a special structure. In this paper, we focus on  the performance of the MCCS with optimized cache placement in  {\bf P0}. In particular, we will analyze the performance gap between the optimized MCCS and the lower bounds of  caching developed in section \ref{sec:lb}. %  in Section \ref{sec:lb}.
\end{Remark}

%%%%%%%%%%%%%%%%%%%%%%%%%%%%%%%%%%%%%%%%%%%%%%%%%%%%%%%%%
\section{Converse Bound for Uncoded Placement}\label{sec:lb}
In this section, we  first introduce a lower bound on the average rate for caching under  any uncoded placement. Then, we develop a popularity-first-based lower bound by restricting to the  popularity-first placement strategies.

Let $\Dc$ be the set of the  distinct file indexes  in demand vector $\dbf$, \ie  $\Dc = \Unique(\dbf)\subseteq \Nc$, where $\Unique(\dbf)$ is to extract the unique elements in $\dbf$.
Also, recall that the users in the leader group $\Uc$  request for  distinct files in $\Dc$.
Thus, we have $|\Dc|=|\Uc|=\tilde{N}(\dbf)$ for a given $\dbf$.
 In the following lemma, we  present the lower bound on the average rate under any uncoded placement.

\vspace{-0.3em}
\begin{lemma}\label{lemma_bnd_2}
For the caching problem described in Section~\ref{sec:model}, the following optimization problem provides a lower bound on the average rate under uncoded placement\\[-1.5em]
\begin{align}
\textrm{\bf P1:}\;\; \min_{\abf}\; \bar R_\text{lb}(\abf)&\triangleq \sum_{{\Dc}\subseteq\Nc} \sum_{\dbf\in\Tc({\Dc})}\prod_{i=1}^{K}p_{d_i} R_\text{lb}({\Dc};\abf)\nn\\ \quad
      \textrm{s.t.} &\quad \eqref{Constraint1.1},\eqref{Constraint2}, \text{~and~} \eqref{Constraint_gt0}\nn
\end{align}\\[-1.5em]
where $\Tc({\Dc})\triangleq \{\dbf: \Unique(\dbf)=\Dc, \ \dbf\in \Nc^K\}$, and $R_\text{lb}({\Dc};\abf)$ is the  lower bound for the distinct file set ${\Dc}$ with the placement vectors $\{\abf_n, n\in{\Dc}\}$, given by\vspace{-0.8em}
\begin{align}\label{R_lb}
 R_\text{lb}(\Dc;\abf)\triangleq\max_{\pi:[|\Dc|]\rightarrow\Dc} \sum_{l=0}^{K-1}\sum_{i=1}^{\tilde{N}(\dbf)}\binom{K-i}{l}a_{\pi(i),l}
\end{align}
where $\pi:[|\Dc|]\rightarrow\Dc$ is any bijective map from $|\Dc|$ to file indexes in    ${\Dc}$.
\end{lemma}\vspace{-0.5em}
\IEEEproof
The result follows  immediately the proof of \cite[Theorem 2]{Sahraei2019TheOptimal} with some slight variations. Details are omitted.
\endIEEEproof
Note that {\bf P1} is a min-max problem. It can be easily converted into an LP problem using  the epigraph form by moving \eqref{R_lb} to the constraints and solved by the standard LP  software.

Recall that the popularity-first placement has been considered in the existing works to simplify the  placement problem for coded caching under nonuniform file popularity. In the following, we develop a popularity-first-based lower bound   for caching, assuming that the  popularity-first placement is used for the uncoded placement.

\begin{lemma}\label{lemma:lbpf}
(Popularity-first-based lower bound) The following  optimization problem provides a lower bound on the average rate for the caching  under any popularity-first  placement\\[-1.4em]
\begin{align}
\textrm{\bf P2:}\; \min_{\abf\in\Qc}\; \bar R_\text{lb}(\abf)&\triangleq \sum_{{\Dc}\subseteq\Nc} \sum_{\dbf\in\Tc({\Dc})}\prod_{i=1}^{K}p_{d_i} R_\text{lb}({\Dc};\abf)\label{equ:converse_obj_pf}\\
\quad      \textrm{s.t.} &\quad \eqref{Constraint1.1},\eqref{Constraint2}, \ \text{and}\  \eqref{Constraint_gt0}\nn
\end{align}%\\[-1.35em]
 where $R_\text{lb}({\Dc};\abf)$, for $\abf\in\Qc$, is simplified to
 \vspace*{-0.5em}
\begin{align}
 R_\text{lb}(\Dc;\abf)\triangleq \sum_{l=0}^{K-1}\sum_{i=1}^{\tilde{N}(\dbf)}\binom{K-i}{l}a_{\phi(i),l},\quad \abf\in\Qc \label{equ:bnd_5}
\end{align}
in which $\phi:[|\Dc|]\rightarrow\Dc$ is a bijective map from $|\Dc|$ to file indexes in    ${\Dc}$, such that the files are sorted in decreasing order over their popularities.
\end{lemma}
Note that by focusing on the set of popularity-first placements  $\Qc$, we remove the ``max" operation in \eqref{R_lb} to arrive at \eqref{equ:bnd_5}.
In the following theorem, we show that for $K=2$ users, such restriction does not lose any optimality.
\begin{theorem}\label{thm:lbequ}
For $K=2$, {\bf P1} and {\bf P2} are equivalent.
\end{theorem}
Theorem \ref{thm:lbequ} indicates that, for nonuniform popularity, the lower bound  for caching under any uncoded placement  in {\bf P1} is attained by the popularity-first placement ({\bf P2}) for $K=2$. For $K>2$, we provide numerical studies in Section \ref{sec:num} to show that the two lower bounds in {\bf P1} and {\bf P2} are generally equal. \vspace{-0.3em}

%%%%%%%%%%%%%%%%%%%%%%%%%%%%%%%%%%%%%%%%%%
%%%%%%%%%%%%%%%%%%%%%%%%%%%%%%%%%%%%%%%%%%
\section{Memory-Rate Tradeoff Characterization  }\label{sec:exact}
In this section, we focus on discussing the gap between the average rate of the optimized MCCS in {\bf P0} and the popularity-first-based lower bound given by {\bf P2}.
Note that the difference between {\bf P0} and {\bf P2} is only in the average rate objective expressions. To show the equivalence of  {\bf P0} and {\bf P2}, it is sufficient to show that $\bar R_\text{MCCS}$ and $\bar R_\text{lb}$ are equal for any $\abf\in\Qc$.

Consider the caching problem with any $N$ files with popularity distribution $\pbf$, and local cache size $M$. To show the tightness of the lower bound, we compare $\bar R_\text{MCCS}$ and $\bar R_\text{lb}$ in the following three possible regions:
\emph{Region 1}: $K=2$; \emph{Region 2}: $K> 2$, $\tilde N(\dbf)=K$ (no  redundant file requests); and  \emph{Region 3}: $K>2$, $\tilde N(\dbf)<K$ (with redundant file requests).
Note that  Region 2 is possible only when  $K\le N$.  For Region 3, there are multiple users  requesting  the same file.
We summarize our results as follows:
\begin{itemize}
  \item For both Regions 1 and 2, we prove that the popularity-first-based lower bound by {\bf P2} is tight, \ie   the optimized MCCS by {\bf P0} attains this lower bound. In particular, in Region 1,   we show the optimality of the MCCS under the popularity-first placement for caching under uncoded placement. Also,  the tight bound reveals that there is no loss of optimality by  zero-padding  in coded messages in the MCCS in both Regions 1 and 2.
\item For Region 3, there may be a gap between the average rate of the optimized MCCS and the popularity-first-based lower bound by {\bf P2}.     It comes from the loss due to zero-padding used during the delivery phase. Nonetheless, the numerical results show that the loss is very small and only appears in limited  scenarios.
\end{itemize}%%%%%%%%%%%%%%%%%%%%%%%
%\vspace*{-.6em}
\subsection{Expression of $\bar R_\text{MCCS}$} \label{sec:achiev}
We first rewrite the expression of $\bar R_\text{MCCS}(\abf)$ in   \eqref{avgR_MCCS} for the MCCS.
Given placement $\abf$, the delivery rate  $R_{\text{MCCS}} (\dbf;\abf)$  in   \eqref{avgR_MCCS} for demand vector $\dbf$ can be rewritten as
\vspace{-0.5em}
\begin{align}
R_{\text{MCCS}} (\dbf;\abf) =\sum_{l=0}^{K-1}\sum_{\Sc\subseteq\Ac^{l+1}, \Sc \cap \Uc\neq\emptyset}\!\!\!\max_{k\in\Sc}a_{d_k,l} \label{equ:converse_achi_obj}
\end{align}%\\[-1em]
where we regroup the terms in $R_{\text{MCCS}} (\dbf;\abf)$ based on the size $|\Sc|$ of the non-redundant groups. Define $\psi\!:\![|\Uc|]\!\to\!\Uc$ as a bijective map to the user indexes in $\Uc$, such that  $p_{d_{\psi(1)}}\ge\ldots\ge p_{d_{\psi(\tilde N(\dbf))}}$. From $\phi:[|\Dc|]\rightarrow\Dc$ defined in Lemma~\ref{lemma:lbpf}, we have  $d_{\psi(i)}=\phi(i), i=1,\ldots,\tilde N(\dbf)$. Since  $\abf\in\Qc$, we have $a_{d_{\psi(1)},l}\geq\ldots\geq a_{d_{\psi(\tilde N(\dbf))},l}$.

 We now partition the coded messages indicated in $R_{\text{MCCS}} (\dbf;\abf)$  into different categories based on the user subsets the messages are corresponding to. Recall that $\Ac^{l+1}$ is the set of all  $\binom{K}{l+1}$ user subsets with size $|\Sc|=l+1$. Among these subsets, there are $\binom{K-1}{l}$ subsets containing  user $\psi(1)$.
We denote the length of the coded message to each of these $\binom{K-1}{l}$ subsets containing user $\psi(1)$ as%\\[-1.5em]
\begin{align}
{\bar a}_{\psi(1),l}\triangleq \max_{k\in\Sc, \Sc\subseteq\Kc, |\Sc|=l+1, \psi(1)\in\Sc \cap \Uc } a_{d_k,l}. \label{equ:ach_obj_L1}
\end{align}\\[-0.8em]
Note that the number of user subsets in $\Ac^{l+1}$ that include $\psi(i)$ but not  $\psi(1),\ldots,\psi(i-1)$ is $\binom{K-i}{l}$. We denote the length of the coded message to each of these subsets as\\[-1.5em]
\begin{align}
\bar{a}_{\psi(i),l}\triangleq\max_{\substack{k\in\Sc, \Sc\subseteq\Kc\backslash\{\psi(1),\ldots,\psi(i-1)\}\\  |\Sc|=l+1,  \psi(i)\in\Sc \cap \Uc } }a_{d_k,l}.\label{equ:ach_obj_L3}
\end{align}%\\[-.em]
Following the above, we can rewrite \eqref{equ:converse_achi_obj} as
\vspace{-0.9em}
\begin{align}\label{equ:achiev_conv_1}
R_{\text{MCCS}}(\dbf;\abf) = \sum_{l=0}^{K-1}\sum_{i=1}^{\tilde{N}(\dbf)}\sum_{\substack{ \Sc\subseteq\Kc\backslash\{\psi(1),\ldots,\psi(i-1)\}\\  |\Sc|=l+1,  \psi(i)\in\Sc \cap \Uc }}\bar{a}_{\psi(i),l}.
\end{align}
We point out that the exact value of $\bar{a}_{\psi(i),l}$ depends on the specific user subset that includes $\psi(i)$ but not $\{\psi(1),\ldots,\psi(i-1)\}$.
Averaging $R_{\text{MCCS}}(\dbf;\abf)$ in \eqref{equ:achiev_conv_1} over $\dbf$, we rewrite
 $\bar R_\text{MCCS}(\abf)$  in \eqref{avgR_MCCS}  as\\[-1.7em]
 \begin{align}\label{equ:avg_R_MCCS}
\bar R_\text{MCCS}(\abf)=\sum_{\Dc\subseteq\Nc}\sum_{\dbf\in \Tc({\Dc})}\prod_{i=1}^{K}p_{d_i} R_{\text{MCCS}}(\dbf;\abf).
\end{align}%\\[-1em]
where $\Tc({\Dc})$ is defined in Lemma~\ref{lemma_bnd_2}. Using this expression, we now can directly compare the minimum average rate in {\bf P0} and {\bf P1} or {\bf P2}.

\vspace*{-1em}
\subsection{Region 1: $K=2$}\label{sec:two-user}
We have the following result on the optimality of the MCCS.\\[-1.5em]

\begin{theorem}\label{thm_K2}
For  the caching problem of   $N$ files with distribution $\pbf$ and local cache $M$, for $K=2$,  the minimum average rate for the optimized MCCS in {\bf P0} attains the lower bound given by {\bf P1}, and the  MCCS is  optimal  for caching with uncoded placement.%\\[-1.5em]
\end{theorem}\vspace{-0.5em}

\IEEEproof We provide a brief outline of the proof. We first show that  the objective functions $\bar R_\text{MCCS}(\abf)$ in {\bf P0} and  $\bar R_\text{lb}(\abf)$  in {\bf P2} are the same. From \eqref{equ:converse_obj_pf} and \eqref{equ:avg_R_MCCS},   we only need to examine $R_\text{lb}({\Dc};\abf)$ and  $R_{\text{MCCS}}(\dbf;\abf)$ to verify this. Next, by   Theorem \ref{thm:lbequ}, it is straightforward  to see that {\bf P0} and {\bf P1} are equivalent with the same minimum objective value.
\endIEEEproof
Theorem~\ref{thm_K2} indicates both the optimality of the popularity-first placement for the MCCS and the optimality of the MCCS under this placement. This tight bound enables us to characterize the exact memory-rate tradeoff under the uncoded placement. Furthermore, it reveals that zero-padding used in  the MCCS  for the coded message incurs no loss of optimality.

\subsection{Region $2$: $K>2$, $\tilde N(\dbf)=K$}\label{sec:peak-load}
In this case, every user requests for a different file, \ie $|{\Dc}|=|\Uc|=\tilde{N}(\dbf)=K$, where we have an implicit assumption  $K\leq N$.
Let $p_{d_i|K}$ denote the conditional probability of file $i$ being requested, given $\tilde{N}(\dbf) = K$. For $\abf\in\Qc$, the lower bound on the average rate  can be expressed as
%\vspace*{-.5em}
\begin{align}\label{equ:peak_lb}
\bar R_\text{lb}(\abf)&=\sum_{{\Dc}\subseteq\Nc^K} \sum_{\dbf\in\Tc({\Dc})}\prod_{i=1}^{K}p_{d_i|K} R_\text{lb}({\Dc};\abf).
\end{align}
Similarly, the average rate for the MCCS is given by%\\[-1.5em]
\begin{align}\label{equ:peak_obj}
\hspace*{-.5em}\bar R_\text{MCCS}(\abf)&\!=\!\sum_{{\Dc}\subseteq\Nc^K} \sum_{\dbf\in\Tc({\Dc})}\prod_{i=1}^{K}p_{d_i|K}R_{\text{MCCS}}(\dbf;\abf).
\end{align}
Note that $R_\text{lb}({\Dc};\abf)$ and $R_{\text{MCCS}}(\dbf)$ are given in \eqref{equ:bnd_5} and \eqref{equ:achiev_conv_1}, respectively.
Examining the rates in \eqref{equ:peak_lb} and \eqref{equ:peak_obj} for $\abf\in\Qc$ in {\bf P0} and {\bf P2}, we have the following result.\\[-1.5em]

\begin{theorem}\label{thm_peak}
For the caching problem of $N$ files with distribution $\pbf$ and local cache $M$, in Region 2,  the optimized MCCS attains the lower bound on average rate with popularity-first based uncoded placement given by {\bf P2}.\end{theorem}

\subsection{Region $3$: $K>2$, $\tilde N(\dbf)<K$}\label{sec:average-load}
This region reflects the  case when there are multiple users request for the same file. In the following, we  show that  in general there may exit a gap between $R_\text{MCCS}(\dbf;\abf)$ and $R_{\text{lb}}({\Dc};\abf)$  for $\abf\in\Qc$.
The main cause of the gap is the zero-padding used in the MCCS.

From  \eqref{equ:bnd_5} and \eqref{equ:achiev_conv_1}, we see that the number of  coded messages need to be sent by the MCCS in $R_\text{MCCS}(\dbf;\abf)$ is the same as that in  $R_\text{lb}({\Dc};\abf)$, which is $\sum_{l=0}^{K-1}\sum_{i=1}^{\tilde N(\dbf)}\binom{K-i}{l}$. The only difference between  $R_\text{MCCS}(\dbf;\abf)$ and $R_{\text{lb}}({\Dc};\abf)$ is the length of the coded messages, \ie $\bar{a}_{\psi(i),l}$ and $a_{\phi(i),l}$.  Thus, we need to examine whether  $\bar{a}_{\psi(i),l}$ is the same as $a_{\phi(i),l}$. For $|\Dc|=|\Uc|=\tilde{N}(\dbf)=1$, all the users request for the same file.
It follows that $\bar{a}_{\psi(1),l}=a_{d_{\psi(1)},l}=a_{\phi(1),l}$, since only one file needs to be delivered.
However, for $1<\tilde{N}(\dbf)<K$,  there exists a possible gap between $\bar{a}_{\psi(i),l}$ and $a_{\phi(i),l}$ caused by zero-padding, as illustrated in the following example.

\emph{Example:} Assume that there are two users request for file $\phi(1)$. We denote them as $\psi(1)$ and $k_{\phi(1)}\notin \Uc$ (\ie one user is from a redundant group).
By the lower bound $R_\text{lb}({\Dc};\abf)$ in \eqref{equ:bnd_5},  for  all $\binom{K-2}{l}$ user subsets that include user $\psi(2)$ but not user $\psi(1)$, the length of  coded messages corresponding to these subsets is $a_{\phi(2),l}$.
However, from $\bar{a}_{\psi(2),l}$ in \eqref{equ:ach_obj_L1}, by zero-padding, the length of coded messages for the user subsets that include users $\psi(2)$ and $k_{\phi(1)}$ but not user $\psi(1)$ is $\bar{a}_{\psi(2),l}=a_{d_{k_{\phi(1)}},l}=a_{\phi(1),l}$.
In this case,  zero-padding results in longer coded messages correspond to the user subsets that include redundant user $k_{\phi(1)}$ but not the leader user $\psi(1)$, since it always zero-pads to the longest subfile.

%%%%%%%%%%%%%%%%%%%%%%%%%%%%%%%%%%%%%%%%%%%%%%%%%%%%%%%%%%%%%%%%%%%%%%%%%%%%%%%%%%%
\begin{Remark}
As discussed above, when there are redundant file requests, zero-padding  the message to the longest subfile may cause a loss of optimality. One possible solution to avoid this is to create as many subfiles of equal sizes as possible during the placement phase. Coincidentally, such an approach has been exploited in~\cite{Sahraei2019TheOptimal} for the case of two files, where a placement scheme is proposed to  create equal subfile length and is shown to be the optimal caching scheme with uncoded placement for two files.\end{Remark}\vspace{-1em}

\section{Numerical Results}\label{sec:num}
\begin{figure}[t]
\centering
\subfigure[$\theta=0.8$.] %first figure
{
        \includegraphics[scale=0.28]{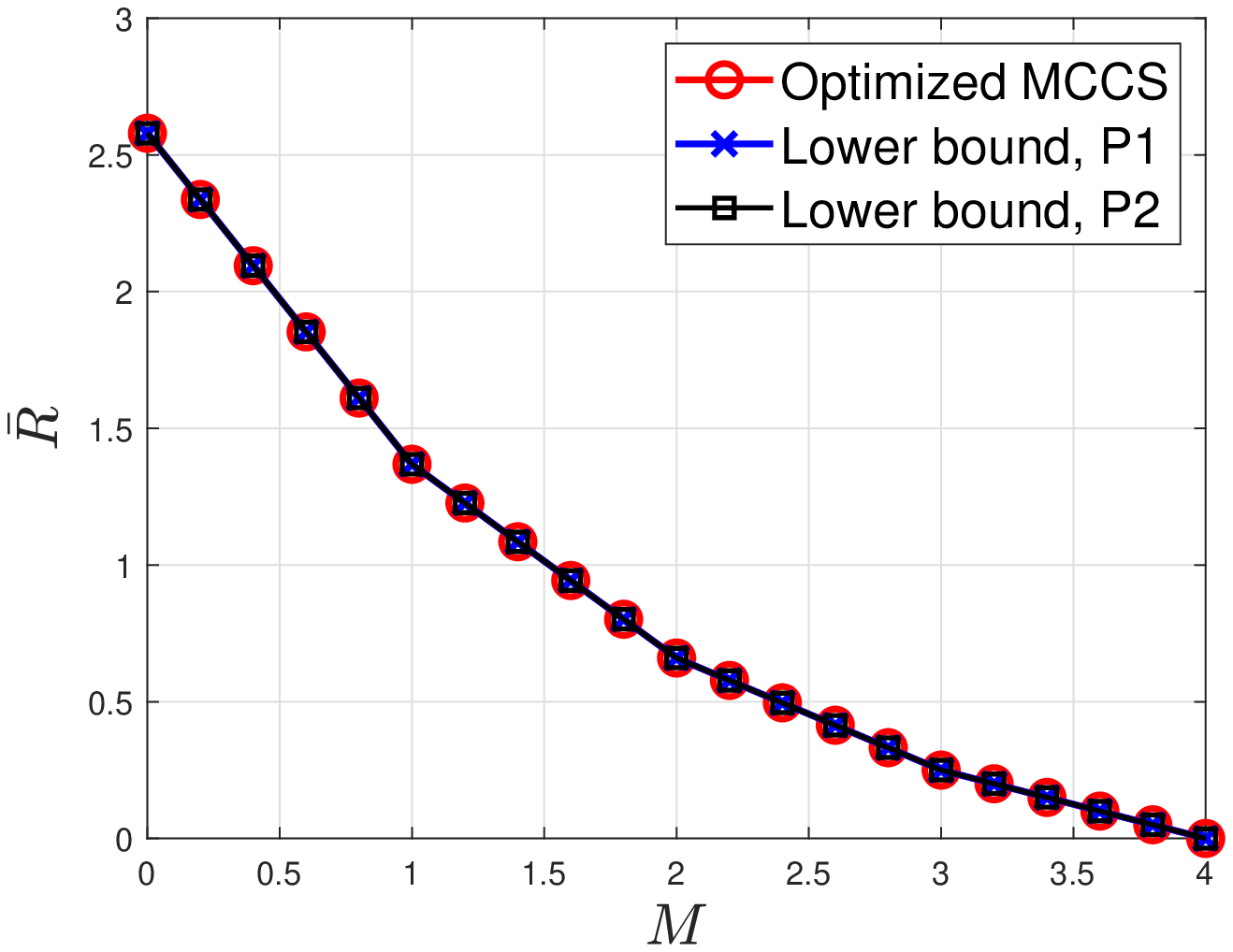}
}
\subfigure[$\theta=1.4$. ] %second figure
{
        \includegraphics[scale=0.28]{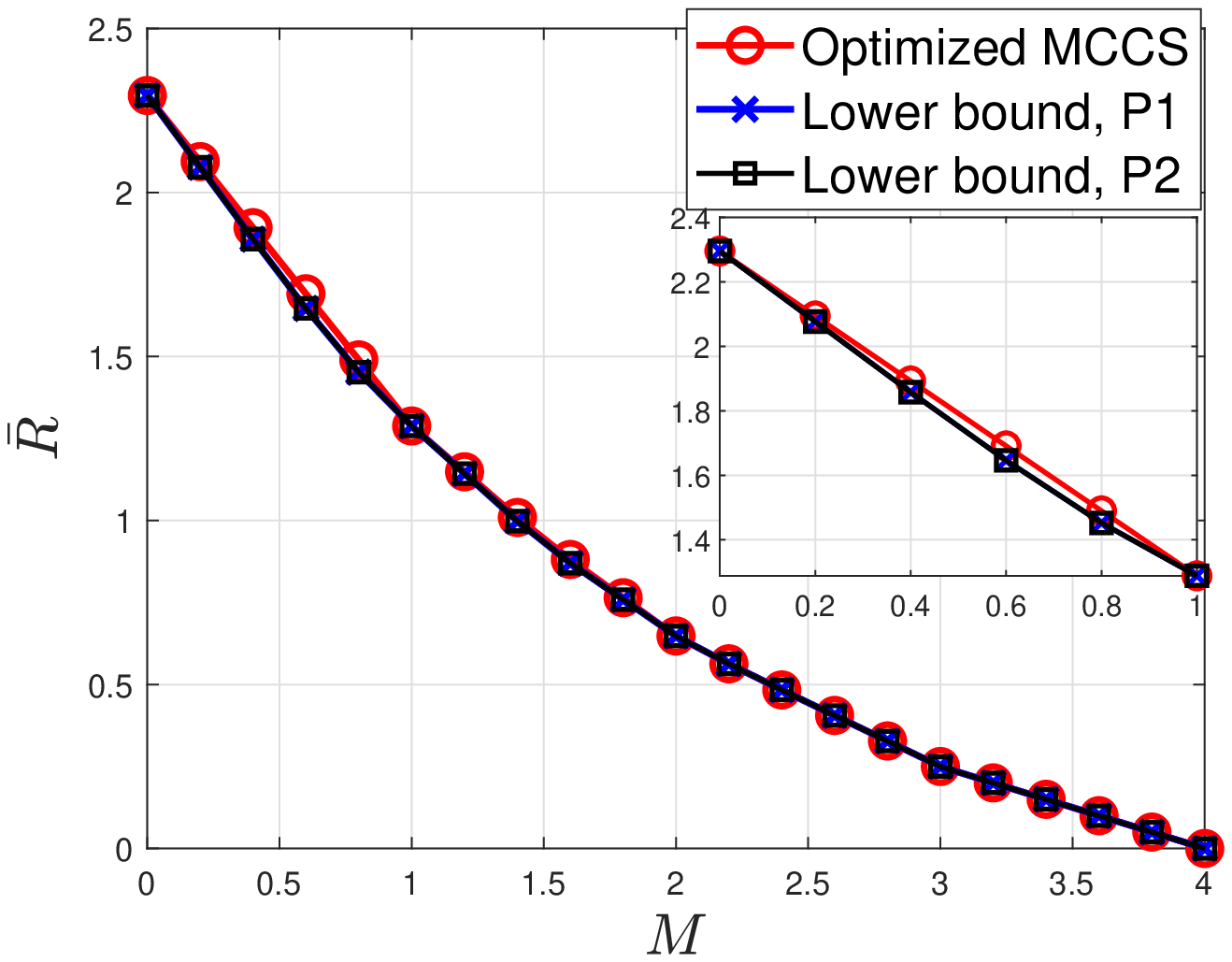}   %
}
\vspace{-0.5em}
\caption{\!Comparison of the optimized MCCS and the lower bounds in {\bf P1} and {\bf P2}: $\bar{R}$ vs. cache size $M$ ($N=4$, $K=4$, Zipf distribution $\theta=0.8$ and $1.4$).}\vspace{-0.5em}
\label{fig:lb-MCCS-theta}  %
\end{figure}
\begin{figure}[tbp]
\centering
\subfigure[$M=0.9$.] %first figure
{
        \includegraphics[scale=0.28]{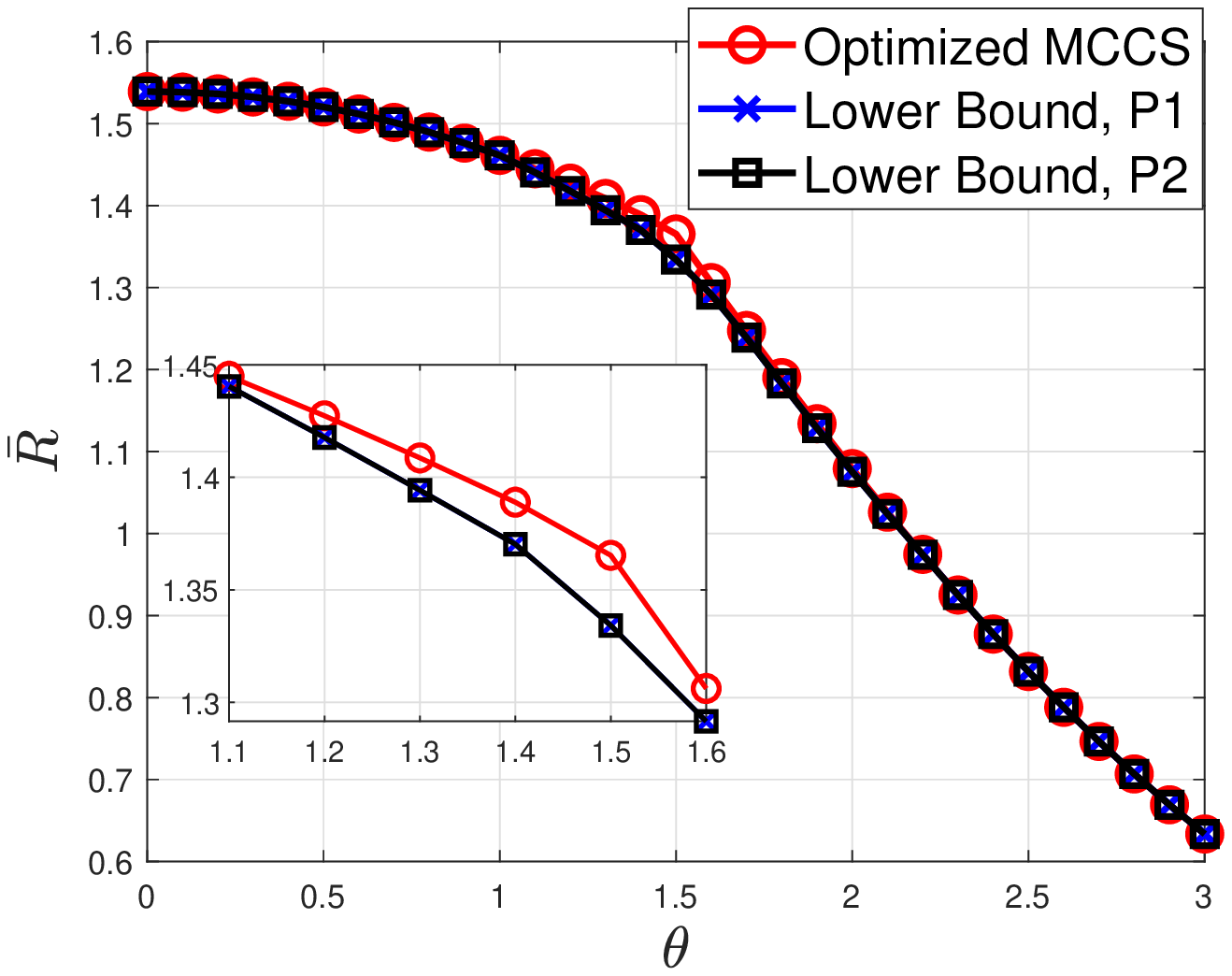}
}
\subfigure[$M=2.1$. ] %second figure
{
        \includegraphics[scale=0.28]{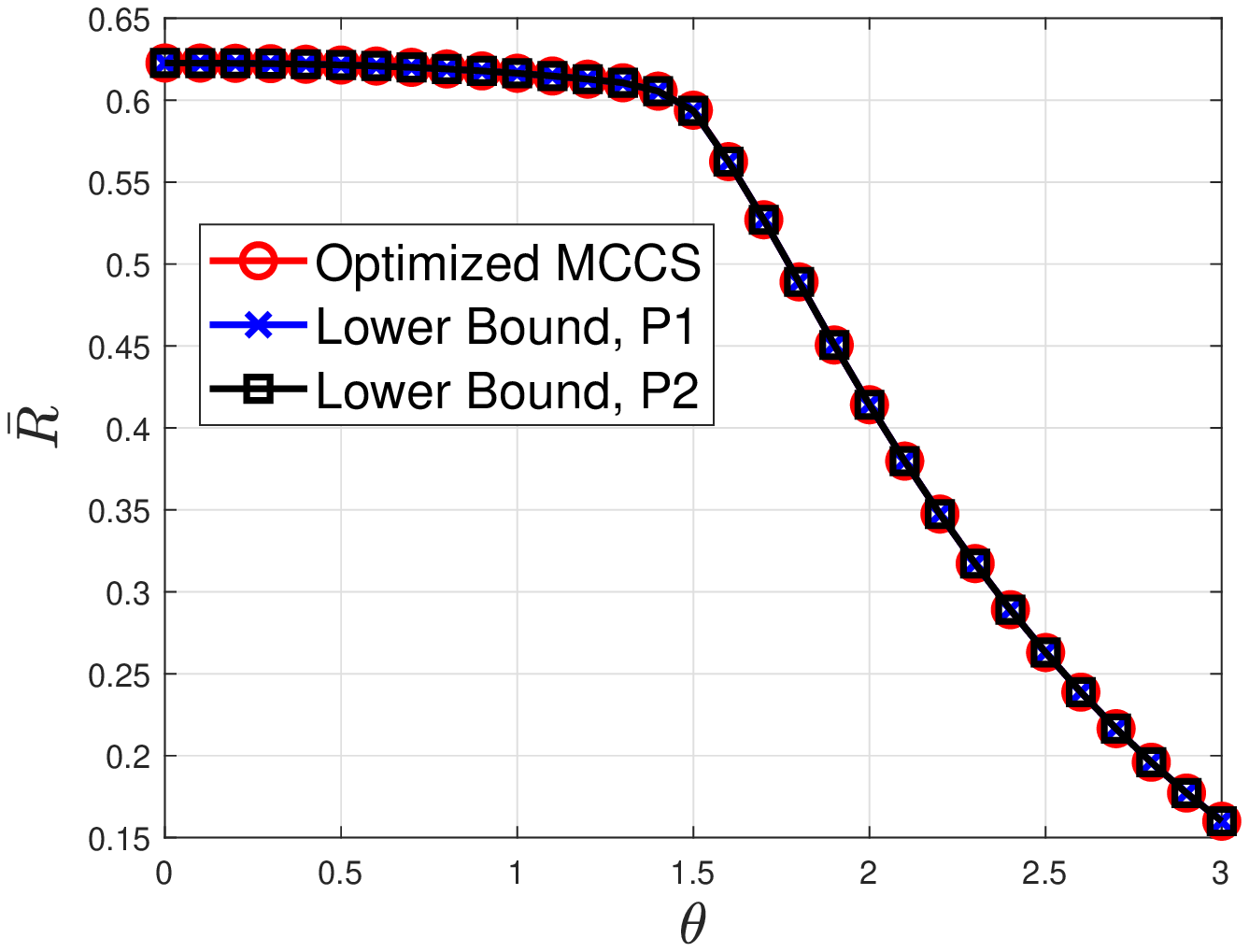}   %
}\vspace{-0.5em}
 \caption{Comparison of the optimized MCCS and the lower bounds in {\bf P1} and {\bf P2}: $\bar{R}$ vs. Zipf distribution $\theta$ ($N=4$, $K=4$, cache size $M=0.9$ and $2.1$).}\vspace{-0.8em}
\label{fig:lb-MCCS-M}  %
\end{figure}
%\vspace{-1em}
In this section, we numerically evaluate the gap between the average rate of the optimized MCCS in {\bf P0} and the proposed lower bounds in {\bf P1} and {\bf P2}. We set $N=4$ files and $K=4$ users. We generate file popularity using Zipf distribution with  $p_{n}={n^{-\theta}}/{\sum_{i=1}^{N}i^{-\theta}}$.
Fig.~\ref{fig:lb-MCCS-theta} shows the average rate vs. cache size $M$ for $\theta=0.8$ and $1.4$.
For both $\theta$ values, the lower bound in {\bf P1} and the popularity-first based lower bound in {\bf P2} are shown to be identical. For $\theta=0.8$, the optimized MCCS attains the lower bound.
For $\theta=1.4$, there is a very small gap between the optimized MCCS and the lower bounds for $M \in (0.1, 1)$. In general, we can see that the gap only exists at smaller $M$ values regardless of $\theta$.
Fig.~\ref{fig:lb-MCCS-M} shows the gap between the optimized MCCS and the lower bounds for $M=0.9$ and $2.1$.
Again, we see the two lower bounds in {\bf P1} and {\bf P2} are identical. Moreover, for $M = 0.9$, we observe a very small gap between the optimized MCCS and the lower bounds for $\theta \in (1.1, 1.6)$.
In general, the  optimized MCCS attains the lower bounds in most of the cases.
The gap only appears in   limited cases and is very small.
%\vspace{-0.5em}

%%%%%%%%%%%%%%%%%%%%%%%%%%
\section{Conclusion}
\label{sec:conclusion}
In this paper, we characterized the memory-rate tradeoff for caching with uncoded placement  under nonuniform file popularity. We considered the\ MCCS with  the optimized cache placement under the popularity-first placement. We provided a general lower bound  and a popularity-first-based lower bound on the average rate. For any $N$ files with popularity $\pbf$ and local cache $M$,  we showed that the optimized MCCS attains the general bower bound for two users and attains the popularity-first-based lower bound for $K>2$ users with no redundant requests.  For $K>2$ users with redundant requests,  there is a possible gap between the MCCS and the lower bounds due to zero-padding. Nonetheless, numerical results show that the gap  is generally very small and only exists in some limited cases.
%\vspace{-0.5em}
\bibliographystyle{IEEEtran}
\bibliography{Yong,IEEEabrv}

\begin{thebibliography}{10}
\providecommand{\url}[1]{#1}
\csname url@samestyle\endcsname
\providecommand{\newblock}{\relax}
\providecommand{\bibinfo}[2]{#2}
\providecommand{\BIBentrySTDinterwordspacing}{\spaceskip=0pt\relax}
\providecommand{\BIBentryALTinterwordstretchfactor}{4}
\providecommand{\BIBentryALTinterwordspacing}{\spaceskip=\fontdimen2\font plus
\BIBentryALTinterwordstretchfactor\fontdimen3\font minus
  \fontdimen4\font\relax}
\providecommand{\BIBforeignlanguage}[2]{{%
\expandafter\ifx\csname l@#1\endcsname\relax
\typeout{** WARNING: IEEEtran.bst: No hyphenation pattern has been}%
\typeout{** loaded for the language `#1'. Using the pattern for}%
\typeout{** the default language instead.}%
\else
\language=\csname l@#1\endcsname
\fi
#2}}
\providecommand{\BIBdecl}{\relax}
\BIBdecl

\bibitem{Maddah-Ali&Niesen:TIT2014}
M.~A. Maddah-Ali and U.~Niesen, ``Fundamental limits of caching,'' \emph{{IEEE}
  Trans. Inform. Theory}, vol.~60, pp. 2856--2867, Mar. 2014.

\bibitem{Niesen&Maddah-Ali:TIT2015}
------, ``Decentralized coded caching attains order-optimal memory-rate
  tradeoff,'' \emph{{IEEE/ACM} Trans. Netw.}, pp. 1029--1040, Aug. 2015.

\bibitem{ji2016fundamental}
M.~Ji, G.~Caire, and A.~F. Molisch, ``Fundamental limits of caching in wireless
  {D2D} networks,'' \emph{{IEEE} Trans. Inform. Theory}, pp. 849--869, Feb.
  2016.

\bibitem{Xu&etal:TIT17}
F.~Xu, M.~Tao, and K.~Liu, ``Fundamental tradeoff between storage and latency
  in cache-aided wireless interference networks,'' \emph{{IEEE} Trans. Inform.
  Theory}, pp. 7464--7491, Jun. 2017.

\bibitem{Yu&Maddah-Ali:TIT2018}
Q.~Yu, M.~A. Maddah-Ali, and A.~S. Avestimehr, ``The exact rate-memory tradeoff
  for caching with uncoded prefetching,'' \emph{{IEEE} Trans. Inform. Theory},
  pp. 1281--1296, Feb. 2018.

\bibitem{yapar2019optimality}
\c{C}. {Yapar}, K.~{Wan}, R.~F. {Schaefer}, and G.~{Caire}, ``On the optimality
  of {D2D} coded caching with uncoded cache placement and one-shot delivery,''
  \emph{{IEEE} Trans. Commun.}, pp. 8179--8192, Dec. 2019.

\bibitem{Wan2016On1}
K.~{Wan}, D.~{Tuninetti}, and P.~{Piantanida}, ``On caching with more users
  than files,'' in \emph{Proc. {IEEE} Int. Symp. on Infor. Theory (ISIT)}, Jul.
  2016.

\bibitem{Wan2016On2}
------, ``On the optimality of uncoded cache placement,'' in \emph{{IEEE}
  Infor. Theory Workshop}, Sep. 2016.

\bibitem{Niesen&Maddah-Ali:TIT2017}
U.~Niesen and M.~A. Maddah-Ali, ``Coded caching with nonuniform demands,''
  \emph{{IEEE} Trans. Inform. Theory}, pp. 1146--1158, Dec. 2017.

\bibitem{Ji&Order:TIT17}
M.~Ji, A.~M. Tulino, J.~Llorca, and G.~Caire, ``Order-optimal rate of caching
  and coded multicasting with random demands,'' \emph{{IEEE} Trans. Inform.
  Theory}, pp. 3923--3949, Apr. 2017.

\bibitem{Zhang&Coded:TIT18}
J.~Zhang, X.~Lin, and X.~Wang, ``Coded caching under arbitrary popularity
  distributions,'' \emph{{IEEE} Trans. Inform. Theory}, pp. 349--366, Nov.
  2018.

\bibitem{Daniel&Yu:TIT19}
A.~M. {Daniel} and W.~{Yu}, ``Optimization of heterogeneous coded caching,''
  \emph{{IEEE} Trans. Inform. Theory}, vol.~66, pp. 1893--1919, Mar. 2020.

\bibitem{Jin&Cui:Arxiv2017}
S.~Jin, Y.~Cui, H.~Liu, and G.~Caire, ``Structural properties of uncoded
  placement optimization for coded delivery,'' \emph{arXiv preprint
  arXiv:1707.07146}, Jul. 2017.

\bibitem{Jin&Cui:Arxiv2018}
------, ``Uncoded placement optimization for coded delivery,'' \emph{arXiv
  preprint arXiv:1709.06462}, Jul. 2018.

\bibitem{Deng&Dong:arxiv2020}
Y.~Deng and M.~Dong, ``Fundamental structure of optimal cache placement for
  coded caching with heterogeneous demands,'' \emph{arXiv preprint
  arXiv:1912.01082}, Apr. 2020.

\bibitem{Yong2020Optimal}
Y.~{Deng} and M.~{Dong}, ``Optimal uncoded placement and file grouping
  structure for improved coded caching under nonuniform popularity,'' in
  \emph{the 18th Int. Symposium on Modeling and Opt. in Mobile, Ad Hoc, and
  Wireless Netw. (WiOPT)}, 2020.

\bibitem{Sahraei2019TheOptimal}
S.~{Sahraei}, P.~{Quinton}, and M.~{Gastpar}, ``The optimal memory-rate
  trade-off for the non-uniform centralized caching problem with two files
  under uncoded placement,'' \emph{{IEEE} Trans. Inform. Theory}, pp.
  7756--7770, Dec. 2019.

\bibitem{Deng&Dong:SPAWC19}
Y.~{Deng} and M.~{Dong}, ``Optimal cache placement for modified coded caching
  with arbitrary cache size,'' in \emph{Proc. {IEEE} Int. Workshop on Signal
  Processing advances in Wireless Commun.(SPAWC)}, Jul. 2019, pp. 1--5.

\bibitem{Saberali&Lampe:TIT20}
S.~A. {Saberali}, L.~{Lampe}, and I.~F. {Blake}, ``Full characterization of
  optimal uncoded placement for the structured clique cover delivery of
  nonuniform demands,'' \emph{{IEEE} Trans. Inform. Theory}, vol.~66, pp.
  633--648, Jan. 2020.

\end{thebibliography}

\end{document}